\documentclass[prb,superscriptaddress,twocolumn,floatfix]{revtex4-2}
\usepackage{balance}
\usepackage{mathptmx}
\usepackage{graphicx} 
\usepackage{lastpage}
\usepackage{placeins}
\usepackage{amssymb}
\usepackage{booktabs}
\usepackage[format=plain,justification=raggedright,singlelinecheck=false,font={stretch=1.125,small,sf},labelfont=bf,labelsep=space]{caption}
\usepackage{float}
\usepackage{fancyhdr}

\usepackage{fnpos}
\usepackage[english]{babel}
\addto{\captionsenglish}{%
  
}
\usepackage{array}
\usepackage{chemformula}
\usepackage{droidsans}
\usepackage{charter}
\usepackage[T1]{fontenc}
\usepackage[usenames,dvipsnames]{xcolor}
\usepackage{setspace}
\usepackage[compact]{titlesec}
\usepackage[hidelinks]{hyperref}
\hypersetup{
  colorlinks=true,
  citecolor=blue,
  filecolor=blue,
  linkcolor=blue,
  urlcolor=blue,
  breaklinks=true
}
\renewcommand{\doi}[1]{\href{https://doi.org/#1}{doi:\nolinkurl{#1}}}
\usepackage{epstopdf}
\usepackage{tikz}
\usetikzlibrary{arrows.meta,positioning,calc,shapes.geometric,fit}

\usepackage{listings}
\definecolor{cream}{RGB}{222,217,201}
\definecolor{accent}{RGB}{31,77,106}
\definecolor{bar1}{RGB}{31,77,106}
\definecolor{bar2}{RGB}{55,105,135}
\definecolor{bar3}{RGB}{90,140,165}
\definecolor{bar4}{RGB}{130,175,195}
\definecolor{bar5}{RGB}{165,200,215}
\definecolor{bar6}{RGB}{190,210,222}
\definecolor{barother}{RGB}{175,165,150}
\definecolor{codegreen}{rgb}{0,0.6,0}
\definecolor{codegray}{rgb}{0.5,0.5,0.5}
\definecolor{codepurple}{rgb}{0.58,0,0.82}
\definecolor{backcolour}{rgb}{0.95,0.95,0.92}
\lstdefinestyle{mystyle}{
    backgroundcolor=\color{backcolour},   
    commentstyle=\color{codegreen},
    keywordstyle=\color{magenta},
    numberstyle=\tiny\color{codegray},
    stringstyle=\color{codepurple},
    basicstyle=\ttfamily\footnotesize,
    breakatwhitespace=false,         
    breaklines=true,                 
    captionpos=b,                    
    keepspaces=true,                 
    numbersep=5pt,                  
    showspaces=false,                
    showstringspaces=false,
    showtabs=false,                  
    tabsize=2
}
  
\begin{document}

\title{Molecular Crystal Structure Prediction from Conditional Flow on the Unit Cells}
\author{Qiang Zhu}
\email{qzhu8@charlotte.edu}
\affiliation{Department of Mechanical Engineering and Engineering Science, University of North Carolina at Charlotte, Charlotte, NC 28223, USA}
\affiliation{North Carolina Battery Complexity, Autonomous Vehicle and Electrification (BATT CAVE) Research Center, Charlotte, NC 28223, USA}

\author{Yihan Weng}
\affiliation{Department of Mechanical Engineering and Engineering Science, University of North Carolina at Charlotte, Charlotte, NC 28223, USA}

\date{\today}
\begin{abstract}
\textbf{Abstract.}
A molecular crystal structure is jointly described by its space group symmetry, unit cell, and the molecular alignment within the asymmetric unit. Concurrently predicting all three variables is a daunting task, as it mixes discrete symmetry choices with a high-dimensional search in the continuous space. To address this challenge, we decouple these variables using a three-step generation process. Specifically, we train a flow model to learn the conditional distribution of invariant lattice descriptors (e.g. direct- and reciprocal-lattice successive minima and Selling scalars) from a molecular graph, a Hall setting, and the number of molecules in the asymmetric unit ($Z'$). Using a two sequential quasi-random sampling processes, we first reconstruct the cell parameters that match the predicted lattice invariants and density requirements, and then conduct a molecular packing search within the give symmetry and unit cell constraint. On 84 single-component systems with $Z' \le 1$, our approach reproduces experimental matches for 83 systems; the remaining failure stems from force-field limitations in preserving the experimental structure. These results demonstrate that learned cell proposals can effectively support crystal structure prediction (CSP) for a given Hall setting and $Z'$, which may be extended to fully blind prediction with variable symmetry and $Z'$ settings in the future.
\end{abstract}

\maketitle
\makeatletter

\vspace{3mm}\noindent

\section{Introduction}
In the context of organic materials research, crystal structure prediction (CSP) aims to answer which periodic packing a given molecule can adopt and be observed under given conditions~\cite{Price2014,Oganov-NRM-2019,zhu2023organic}. The question matters because many organic molecules can crystallize in multiple forms, and their polymorphs may differ in solubility, stability, mechanical behaviour, and optical or electronic properties. Progress on this problem has been measured by the blind tests organized periodically by the Cambridge Crystallographic Data Centre~\cite{Reilly2016,Hunnisett2024}. A CSP method that is both reliable and inexpensive would extend this capability from case-by-case studies to routine use such as screening polymorph risk early in drug development and designing new organic materials.

Over the past two decades, CSP methods have made steady progress in both structure generation and energy ranking. A search stage generates a large population of trial crystals by sampling space groups, unit cells, and molecular positions, orientations, and conformations, using random or quasi-random generation, evolutionary and genetic algorithms, or population-based sampling under symmetry constraints~\cite{Oganov2011,zhu2012,Curtis2018,Fredericks2021,ZhuHattori2025}. A ranking stage then orders the candidates by lattice energy at successively higher levels of theory~\cite{Neumann2008,Price2014}. This strategy has become increasingly successful, but its cost remains high: because the search is blind to which cells a molecule favors, tens of thousands to millions of trial structures must be generated and relaxed before the landscape converges.

To make CSP scalable to large numbers of molecules, a model is needed that can propose plausible unit cells and molecular placements far faster than blind search. AI generative models offer such a route. Originally developed for image synthesis, they learn to draw samples from a target distribution and can then propose new candidates at negligible cost. Their success in materials discovery was first demonstrated for inorganic solids with small unit cells, using diffusion, flow-matching, and transformer architectures~\cite{Jiao2023,zeni2025generative,miller2024flowmm}. Similar ideas have since been actively explored for molecular crystals, with modifications that preserve molecular connectivity during generation~\cite{Lo2026Clari,zeng2026molcrystalflow,jin2026oxtal,guo2025assembleflow,vahediahmar2026orgflow}. These models are fast and promising, but their accuracy is not yet sufficient to deliver the final answer on their own. 

A molecular crystal is specified by three ingredients: a space-group setting, a unit cell, and the positions, orientations, and internal degrees of freedom of the molecules in the asymmetric unit. Searching jointly over these discrete and continuous variables is challenging. However, the three ingredients are not equally hard to predict, and they need not be treated by the same tool. The space-group setting is a discrete choice from a short list, so it can be enumerated or drawn from its empirical distribution rather than searched. The unit cell is continuous but low-dimensional, and its statistics are strongly constrained by the molecule. Both are therefore natural targets for a statistical model trained on the hundreds of thousands of known structures. The molecular placement, by contrast, is high-dimensional, depends on intermolecular interactions that a database-level model captures only coarsely, and is decided by energy differences of a few kJ/mol. It is best left to rigorous physics-based sampling and relaxation. Our central hypothesis is thus that CSP benefits from a hybrid decomposition of two sequential tasks: (i) a generative model handles the symmetry-conditioned unit cell, where the data are plentiful and the tolerance is loose, and (ii) a quasi-random search handles the continuous packing variables, where accuracy matters most. In this work we take the first step by conditioning on the Hall setting and $Z'$ and learning the distribution of unit cells; once a setting and a cell are fixed, quasi-random search can focus on the remaining packing variables~\cite{ZhuHattori2025}.

In this work, we present CellNet, a conditional flow model that learns the distribution of unit cells for a given molecule, Hall setting, and $Z'$, together with a two-stage quasi-random search that turns its proposals into relaxed crystal structures. Section~\ref{sec:method} introduces its training data, the basis-invariant lattice representation, the flow model and two search stages on unit cells and molecular packing. Section~\ref{sec:results} assesses structural coverage and computational cost on 84 single-component systems with $Z' \le 1$, and uses controlled comparisons on identical flow draws to isolate the effect of the frozen-lattice fallback. Section~\ref{sec:discussion} discusses the remaining failures, the role of the energy model, and the extensions needed for fully blind prediction with variable symmetry and $Z'$.

\section{Methods}\label{sec:method}

\subsection{Training data}

We assembled the training dataset from the organic crystal structures compiled by Taniguchi and Fukasawa \cite{TaniguchiFukasawa2025}. The source collection contains more than 170\,000 non-polymeric, solvent-free structures with $R<10\%$ and $Z'=1$. Using less restrictive filtering criteria than the original study, we retained 170\,278 structures and divided them into training and test sets containing approximately 80\% (135\,837 structures) and 20\% (34\,441 structures) of the data, respectively. 

Unlike that the inorganic crystal that are mostly expressed in standard space group settings, many molecular crystals are more conveniently reported in non-standard settings, such as $P2_1/n$ and $Pn$. To distinguish these settings, we use the 530 tabulated Hall settings rather than the 230 space-group types alone. Hall symbols explicitly specify the symmetry generators and origin, distinguishing alternative axis and origin choices within the same space-group type~\cite{Hall1981}.

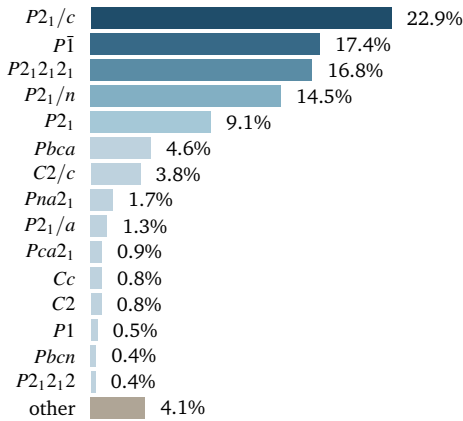
\begin{figure}[t]
\resizebox{0.35\textwidth}{!}{%
\begin{tikzpicture}[x=0.162cm,y=1cm]
  \fill[bar1] (0,4.80) rectangle (22.89,5.06);
  \node[anchor=east,font=\scriptsize] at (-0.4,4.93) {$P2_1/c$};
  \node[anchor=west,font=\scriptsize] at (23.29,4.93) {22.9\%};
  \fill[bar2] (0,4.48) rectangle (17.42,4.74);
  \node[anchor=east,font=\scriptsize] at (-0.4,4.61) {$P\bar{1}$};
  \node[anchor=west,font=\scriptsize] at (17.82,4.61) {17.4\%};
  \fill[bar3] (0,4.16) rectangle (16.84,4.42);
  \node[anchor=east,font=\scriptsize] at (-0.4,4.29) {$P2_12_12_1$};
  \node[anchor=west,font=\scriptsize] at (17.24,4.29) {16.8\%};
  \fill[bar4] (0,3.84) rectangle (14.47,4.10);
  \node[anchor=east,font=\scriptsize] at (-0.4,3.97) {$P2_1/n$};
  \node[anchor=west,font=\scriptsize] at (14.87,3.97) {14.5\%};
  \fill[bar5] (0,3.52) rectangle (9.15,3.78);
  \node[anchor=east,font=\scriptsize] at (-0.4,3.65) {$P2_1$};
  \node[anchor=west,font=\scriptsize] at (9.55,3.65) {9.1\%};
  \fill[bar6] (0,3.20) rectangle (4.58,3.46);
  \node[anchor=east,font=\scriptsize] at (-0.4,3.33) {$Pbca$};
  \node[anchor=west,font=\scriptsize] at (4.98,3.33) {4.6\%};
  \fill[bar6] (0,2.88) rectangle (3.78,3.14);
  \node[anchor=east,font=\scriptsize] at (-0.4,3.01) {$C2/c$};
  \node[anchor=west,font=\scriptsize] at (4.18,3.01) {3.8\%};
  \fill[bar6] (0,2.56) rectangle (1.69,2.82);
  \node[anchor=east,font=\scriptsize] at (-0.4,2.69) {$Pna2_1$};
  \node[anchor=west,font=\scriptsize] at (2.09,2.69) {1.7\%};
  \fill[bar6] (0,2.24) rectangle (1.25,2.50);
  \node[anchor=east,font=\scriptsize] at (-0.4,2.37) {$P2_1/a$};
  \node[anchor=west,font=\scriptsize] at (1.65,2.37) {1.3\%};
  \fill[bar6] (0,1.92) rectangle (0.85,2.18);
  \node[anchor=east,font=\scriptsize] at (-0.4,2.05) {$Pca2_1$};
  \node[anchor=west,font=\scriptsize] at (1.25,2.05) {0.9\%};
  \fill[bar6] (0,1.60) rectangle (0.84,1.86);
  \node[anchor=east,font=\scriptsize] at (-0.4,1.73) {$Cc$};
  \node[anchor=west,font=\scriptsize] at (1.24,1.73) {0.8\%};
  \fill[bar6] (0,1.28) rectangle (0.82,1.54);
  \node[anchor=east,font=\scriptsize] at (-0.4,1.41) {$C2$};
  \node[anchor=west,font=\scriptsize] at (1.22,1.41) {0.8\%};
  \fill[bar6] (0,0.96) rectangle (0.51,1.22);
  \node[anchor=east,font=\scriptsize] at (-0.4,1.09) {$P1$};
  \node[anchor=west,font=\scriptsize] at (0.91,1.09) {0.5\%};
  \fill[bar6] (0,0.64) rectangle (0.41,0.90);
  \node[anchor=east,font=\scriptsize] at (-0.4,0.77) {$Pbcn$};
  \node[anchor=west,font=\scriptsize] at (0.81,0.77) {0.4\%};
  \fill[bar6] (0,0.32) rectangle (0.36,0.58);
  \node[anchor=east,font=\scriptsize] at (-0.4,0.45) {$P2_12_12$};
  \node[anchor=west,font=\scriptsize] at (0.76,0.45) {0.4\%};
  \fill[barother] (0,0.00) rectangle (4.14,0.26);
  \node[anchor=east,font=\scriptsize] at (-0.4,0.13) {other};
  \node[anchor=west,font=\scriptsize] at (4.54,0.13) {4.1\%};
\end{tikzpicture}}
\caption{Distribution of Hall-setting symbols in the 170\,278-structure CellNet extract of SPaDe-CSP (training and test sets combined). }
\label{fig:hall}
\end{figure}

The dataset spans 105 Hall settings, with a strongly imbalanced distribution (Fig.~\ref{fig:hall}). The five most common settings---$P2_1/c$, $P\bar{1}$, $P2_12_12_1$, $P2_1/n$, and $P2_1$---account for 80.8\% of the dataset. The 15 most common settings account for 95.9\%, leaving 4.1\% for all remaining settings. Monoclinic, orthorhombic, and triclinic structures account for 54.4\%, 25.9\%, and 17.9\%, respectively, while tetragonal, trigonal, hexagonal, and cubic structures together account for 1.8\%.

\subsection{Lattice representation}
The unit cell of a molecular crystal is described by three edge lengths and three angles. These conventional parameters, $(a,b,c,\alpha,\beta,\gamma)$, depend on the chosen basis: a unimodular transformation can change the cell edges and angles without changing the lattice. A suitable representation for learning should therefore provide consistent targets for equivalent descriptions of the same lattice while retaining enough metric information to reconstruct a cell. We combine two complementary sets of descriptors as follows.

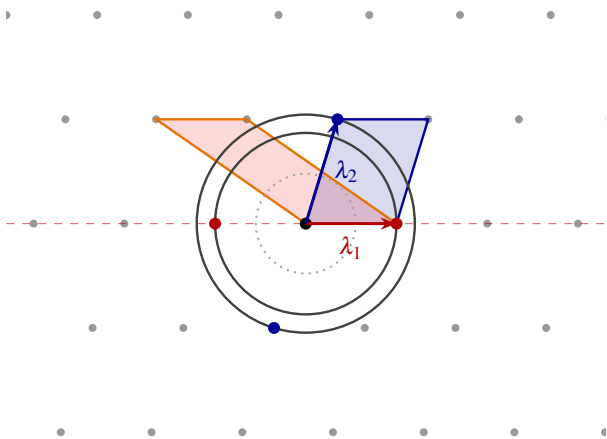
\begin{figure}[htbp]
\centering
\begin{tikzpicture}[
  scale=1.2,
  lat/.style={circle,fill=black!40,inner sep=0pt,minimum size=3pt},
  origin/.style={circle,fill=black,inner sep=0pt,minimum size=4.5pt},
  hit1/.style={circle,fill=red!70!black,inner sep=0pt,minimum size=4.5pt},
  hit2/.style={circle,fill=blue!60!black,inner sep=0pt,minimum size=4.5pt},
  vec1/.style={-{Stealth[length=6pt]},red!70!black,line width=1.1pt},
  vec2/.style={-{Stealth[length=6pt]},blue!60!black,line width=1.1pt},
  cellA/.style={blue!60!black,line width=0.9pt,fill=blue!60!black,fill opacity=0.15},
  cellB/.style={orange!90!black,line width=0.9pt,fill=red!90!black,fill opacity=0.15},
]
\def\xlim{3.3}\def\ylim{2.5}
\begin{scope}
  \clip (-\xlim,-\ylim) rectangle (\xlim,\ylim);
  \foreach \i in {-5,...,5}{\foreach \j in {-3,...,3}{%
    \pgfmathsetmacro{\px}{\i+0.35*\j}\pgfmathsetmacro{\py}{1.15*\j}%
    \node[lat] at (\px,\py) {};}}
  \draw[cellA] (0,0) -- (1,0) -- (1.35,1.15) -- (0.35,1.15) -- cycle;
  \draw[cellB] (0,0) -- (1,0) -- (-0.65,1.15) -- (-1.65,1.15) -- cycle;
  \draw[red!70!black,dashed,thin,opacity=.6] (-\xlim,0) -- (\xlim,0);
  \draw[black!35,dotted,line width=0.8pt] (0,0) circle (0.55);
  \draw[black!75,line width=0.9pt] (0,0) circle (1.0);
  \draw[black!75,line width=0.9pt] (0,0) circle (1.202);
  \node[origin] at (0,0) {};
  \node[hit1] at (1,0) {}; \node[hit1] at (-1,0) {};
  \node[hit2] at (0.35,1.15) {}; \node[hit2] at (-0.35,-1.15) {};
  \draw[vec1] (0,0) -- (1,0)       node[midway,below=1pt,text=red!70!black,font=\small] {$\lambda_1$};
  \draw[vec2] (0,0) -- (0.35,1.15) node[midway,right=1pt,text=blue!60!black,font=\small] {$\lambda_2$};
\end{scope}
\end{tikzpicture}
\caption{Successive minima in two dimensions. A circle centered at a lattice point expands until it reaches a nearest neighboring point (red), defining $\lambda_1$. The radius at which the enclosed lattice vectors span two independent directions defines $\lambda_2$ (blue). The shaded red and blue cells have different edge lengths and angles but describe the same lattice and therefore share the same successive minima.}
\label{fig:minima}
\end{figure}

\vspace{2mm}
\noindent
\textbf{1. Minkowski successive minima~\cite{Cassels1997}.}
The successive minima describe the radii at which an expanding sphere centred at a lattice point encloses independent lattice vectors (see Fig.~\ref{fig:minima}). The first minimum, $\lambda_1$, is the smallest radius at which the sphere encloses a nonzero lattice vector. The second, $\lambda_2$, is the smallest radius at which the enclosed lattice vectors span two independent directions. The third, $\lambda_3$, is the smallest radius at which they span all three dimensions. Formally,
\begin{equation}
\lambda_k(L)=\min\bigl\{r:\ \dim\mathrm{span}\bigl(L\cap\overline{B}(0,r)\bigr)\ge k\bigr\},
\qquad k=1,2,3,
\end{equation}
where $\overline{B}(0,r)$ is the closed ball of radius $r$ centered at the origin. The ordered triple $(\lambda_1,~\lambda_2,~\lambda_3)$ satisfies $\lambda_1\le\lambda_2\le\lambda_3$ and is unchanged by a change of lattice basis. Thus, the two cells in Fig.~\ref{fig:minima} have identical $\lambda$ values despite their different edges and angles.

Applying the same construction to the reciprocal lattice $L^\ast$ defines $\lambda_k^\ast=\lambda_k(L^\ast)$ and the triple $\lambda^\ast=(\lambda_1^\ast,~\lambda_2^\ast,~\lambda_3^\ast)$. The direct and reciprocal minima have units of \AA\ and \AA$^{-1}$, respectively; their logarithms are used as learning targets. Reciprocal minima provide complementary information about the lattice geometry, making the pair $(\lambda,\lambda^\ast)$ more discriminative than either triple alone.

\vspace{2mm}
\noindent
\textbf{2. Delaunay--Selling scalars~\cite{Andrews1988}.}
Starting from a primitive basis $\mathbf{a},\mathbf{b},\mathbf{c}$, we introduce a fourth vector, $\mathbf{d}=-(\mathbf{a}+\mathbf{b}+\mathbf{c})$, so that the four vectors form a superbase whose sum is zero. The six Selling scalars are their pairwise inner products,
$S=\{\mathbf{b}\!\cdot\!\mathbf{c},\ \mathbf{a}\!\cdot\!\mathbf{c},\
\mathbf{a}\!\cdot\!\mathbf{b},\ \mathbf{a}\!\cdot\!\mathbf{d},\
\mathbf{b}\!\cdot\!\mathbf{d},\ \mathbf{c}\!\cdot\!\mathbf{d}\}$,
with units of \AA$^2$. These six scalars encode the metric tensor: the cross terms appear explicitly, and the squared vector lengths follow from $\mathbf{a}+\mathbf{b}+\mathbf{c}+\mathbf{d}=0$. Delaunay reduction produces an obtuse superbase for which all six inner products are non-positive; equivalently, each pair of vectors forms an angle of at least $90^\circ$~\cite{Delaunay1933}. The resulting reduced description removes arbitrary basis choices, subject to permutations of the superbase labels. Within a fixed reduced-superbase representation, the scalars vary smoothly with the metric, providing a compact target for learning lattice geometry.

Table~\ref{tab:selling} illustrates the effect of a basis change for a monoclinic lattice. Replacing $\mathbf{a}$ with $\mathbf{a}'=\mathbf{a}-\mathbf{c}$ increases the corresponding cell edge by more than $5$~\AA\ and changes $\beta$ by more than $40^\circ$. The direct- and reciprocal-lattice successive minima remain unchanged, as do the reduced Selling scalars after consistent ordering of the superbase labels.

\begin{table}[htbp]
\centering
\caption{Two descriptions of one monoclinic lattice.}
\label{tab:selling}
\begin{tabular}{@{}lcc@{}}
\toprule
 & $a,\,b,\,c$ (\AA) & $\alpha,\,\beta,\,\gamma$ ($^\circ$) \\
\midrule
Cell A & 11.23, ~6.54, 11.37 & 90.0, ~95.7, 90.0 \\
Cell B & 16.76, ~6.54, 11.37 & 90.0, 138.2, 90.0 \\
\midrule
$\lambda$ (\AA) & \multicolumn{2}{c}{$(6.54,\ 11.23,\ 11.37)$} \\
$\lambda^\ast$ (\AA$^{-1}$) & \multicolumn{2}{c}{$(0.0884,\ 0.0895,\ 0.1529)$} \\
$S$ (\AA$^2$) & \multicolumn{2}{c}{$(0,\ -12.7,\ 0,\ -113.4,\ -42.8,\ -116.6)$} \\
\bottomrule
\end{tabular}
\end{table}

These descriptors serve complementary roles: the successive minima ensure the reproducibility of unit cell, while the Selling scalars encode the reduced metric and its overall smoothness. Their combination provides geometric targets for both learning and cell reconstruction.

\subsection{The CellNet solution}

CellNet generates plausible unit cells and molecular packings conditioned on a molecular SMILES string, a Hall setting $H$, and the number of molecules in the asymmetric unit, $Z'$. It separates cell generation from symmetry-constrained packing search, using learned unit-cell proposals to guide the search for molecular arrangements.

The workflow comprises three stages (Fig.~\ref{fig:pipeline}). In \textbf{Stage a}, a conditional flow model samples lattice descriptors from the molecular and symmetry inputs, and a separate head predicts density. In \textbf{Stage b}, lattice quasi-random search (QRS) reconstructs Hall-compatible cells and selects a diverse set of candidates. In \textbf{Stage c}, molecular packing QRS places molecules in these cells, relaxes the resulting structures, and ranks them by energy. Experimental structures are used to evaluate coverage and, in the benchmark, to control search termination and the frozen-lattice fallback. They are not used to generate or select cells.

\subsubsection{Conditional flow model}

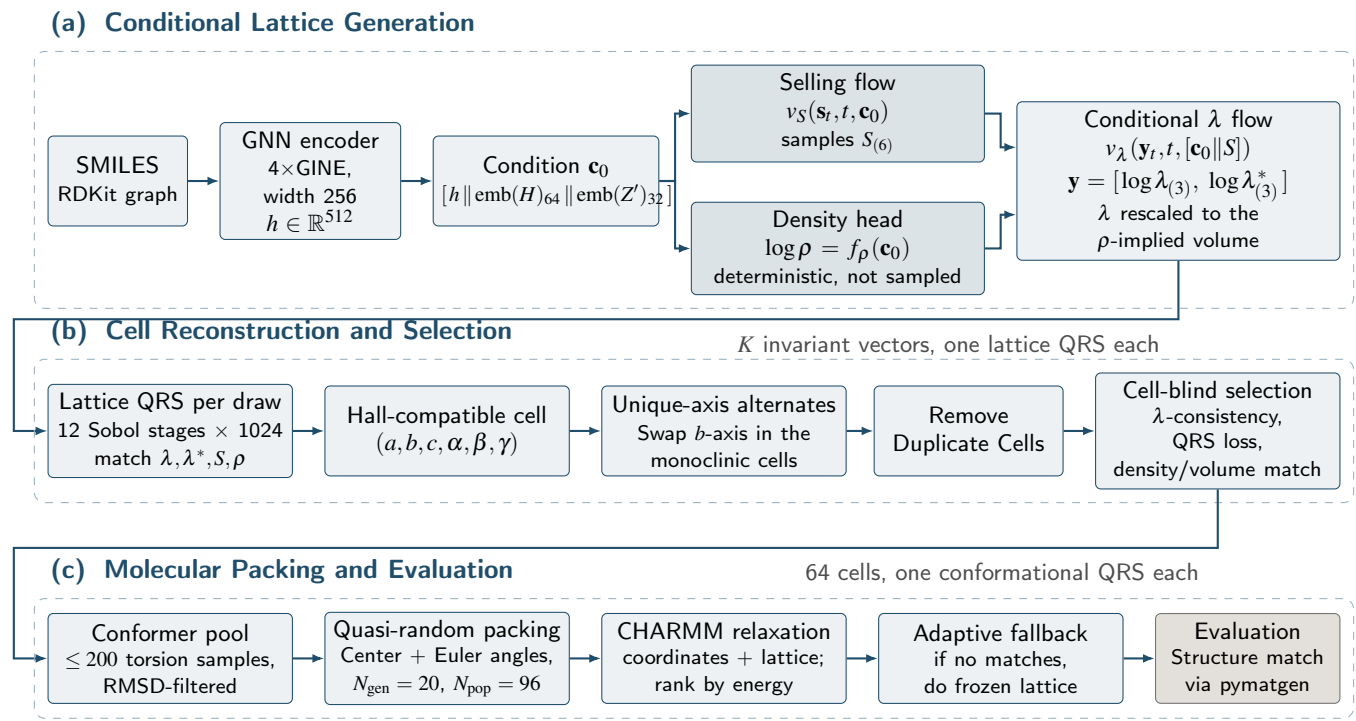
\begin{figure*}[htbp]
\centering
\resizebox{0.99\textwidth}{!}{%
\begin{tikzpicture}[
  font=\sffamily\small,
  box/.style={draw=accent,rounded corners=2pt,align=center,inner sep=3pt,fill=accent!8,text width=3.05cm,minimum height=1.2cm},
  wide/.style={box,text width=4.1cm},
  head/.style={box,fill=accent!16},
  note/.style={font=\sffamily\small,text=black!70,align=center},
  lab/.style={font=\sffamily\bfseries\normalsize,text=accent,anchor=west},
  arr/.style={-{Latex[length=1.6mm]},thick,accent},
  grp/.style={draw=accent!45,dashed,rounded corners=4pt,inner sep=5pt}
]
\node[box,text width=1.65cm] (smi) {SMILES\\{\footnotesize RDKit graph}};
\node[box,right=0.42cm of smi,text width=2.2cm] (gnn) {GNN encoder\\{\footnotesize 4$\times$GINE, width 256}\\$h\in\mathbb{R}^{512}$};
\node[box,right=0.42cm of gnn,text width=2.8cm] (cond) {Condition $\mathbf{c}_0$\\{\footnotesize  $[\,h\,\|\,\mathrm{emb}(H)_{64}\,\|\,\mathrm{emb}(Z')_{32}\,]$}};
\node[head,right=0.42cm of cond,yshift=0.9cm,text width=3.7cm] (sell) {Selling flow\\$v_S(\mathbf{s}_t,t,\mathbf{c}_0)$\\{\footnotesize samples $S_{(6)}$}};
\node[head,right=0.42cm of cond,yshift=-0.9cm,text width=3.7cm] (rho) {Density head\\$\log\rho=f_\rho(\mathbf{c}_0)$\\{\footnotesize deterministic, not sampled}};
\node[wide,right=0.42cm of sell,yshift=-0.9cm,minimum height=2.0cm] (flow) {Conditional $\lambda$ flow\\$v_\lambda(\mathbf{y}_t,t,[\mathbf{c}_0\|S])$\\[2pt]$\mathbf{y}=[\,\log\lambda_{(3)},\;\log\lambda^\ast_{(3)}\,]$\\{\footnotesize $\lambda$ rescaled to the $\rho$-implied volume}};
\draw[arr] (smi) -- (gnn);
\draw[arr] (gnn) -- (cond);
\draw[arr] (cond.east) -- ++(0.21,0) |- (sell.west);
\draw[arr] (cond.east) -- ++(0.21,0) |- (rho.west);
\draw[arr] (sell.east) -- ++(0.21,0) |- ([yshift=0.45cm]flow.west);
\draw[arr] (rho.east) -- ++(0.21,0) |- ([yshift=-0.45cm]flow.west);
\node[grp,fit=(smi)(gnn)(cond)(sell)(rho)(flow)] (g1) {};
\node[lab,anchor=south west] at ([xshift=0.1cm,yshift=0.05cm]g1.north west) {(a)\; Conditional Lattice Generation};

\node[box,anchor=north west] (lqrs) at ([yshift=-2.1cm]smi.south west) {Lattice QRS per draw\\{\footnotesize 12 Sobol stages $\times$ 1024}\\{\footnotesize match $\lambda,\lambda^\ast,S,\rho$}};
\node[box,right=0.42cm of lqrs] (cell) {Hall-compatible cell\\$(a,b,c,\alpha,\beta,\gamma)$\\
};
\node[box,right=0.42cm of cell] (alt) {Unique-axis alternates\\{\footnotesize Swap $b$-axis in the monoclinic cells}};
\node[box,right=0.36cm of alt,text width=2.31cm] (dedup) {Remove\\Duplicate Cells};
\node[box,right=0.42cm of dedup] (sel) {Cell-blind selection \\\footnotesize{$\lambda$-consistency, QRS loss,\\density/volume match}};
\draw[arr] (lqrs) -- (cell);
\draw[arr] (cell) -- (alt);
\draw[arr] (alt) -- (dedup);
\draw[arr] (dedup) -- (sel);
\draw[arr] (flow.south) -- ++(0,-0.78) -| ([xshift=-0.45cm]lqrs.west) -- (lqrs.west);
\node[note,anchor=north east] at ([xshift=-0.12cm,yshift=-0.86cm]flow.south) {$K$ invariant vectors, one lattice QRS each};
\node[grp,fit=(lqrs)(cell)(alt)(dedup)(sel)] (g2) {};
\node[lab,anchor=south west] at ([xshift=0.1cm,yshift=0.05cm]g2.north west) {(b)\; Cell Reconstruction and Selection};

\node[box,below=1.8cm of lqrs] (pool) {Conformer pool\\{\footnotesize $\le 200$ torsion samples,\\RMSD-filtered}};
\node[box,right=0.42cm of pool] (pack) {Quasi-random packing\\{\footnotesize Center $+$ Euler angles,\\$N_{\mathrm{gen}}=20$, $N_{\mathrm{pop}}=96$}};
\node[box,right=0.42cm of pack] (ff) {CHARMM relaxation\\{\footnotesize coordinates $+$ lattice;\\rank by energy}};
\node[box,right=0.42cm of ff] (fb) {Adaptive fallback\\{\footnotesize if no matches,\\do frozen lattice}};
\node[box,right=0.42cm of fb,text width=2.25cm,minimum height=1.0cm,fill=barother!35,draw=black!55] (eval) {Evaluation \\{\footnotesize Structure match \\via pymatgen}};
\draw[arr] (pool) -- (pack);
\draw[arr] (pack) -- (ff);
\draw[arr] (ff) -- (fb);
\draw[arr] (fb) -- (eval);
\draw[arr] (sel.south) -- ++(0,-0.78) -| ([xshift=-0.45cm]pool.west) -- (pool.west);
\node[note,anchor=north east] at ([xshift=-0.12cm,yshift=-0.86cm]sel.south) {64 cells, one conformational QRS each};
\node[grp,fit=(pool)(pack)(ff)(fb)(eval)] (g3) {};
\node[lab,anchor=south west] at ([xshift=0.1cm,yshift=0.05cm]g3.north west) {(c)\; Molecular Packing and Evaluation};
\end{tikzpicture}}
\caption{The CellNet pipeline. (a) A molecular graph and learned Hall and $Z'$ embeddings form the condition $\mathbf{c}_0$. A Selling flow samples the six Selling scalars $S$ from $\mathbf{c}_0$, a second flow samples $[\log\lambda,\log\lambda^\ast]$ given $[\mathbf{c}_0\|S]$ (Fig.~\ref{fig:flow}), and a deterministic head predicts $\log\rho$, to whose implied volume $\lambda$ is rescaled; $K=96$ independent draws are taken. (b) Each draw is inverted to a Hall-compatible cell by quasi-random search on the invariants. (c) PyXtal packs the molecule in each cell and OpenFF/CHARMM relaxes coordinates and lattice for further evaluation.}
\label{fig:pipeline}
\end{figure*}

A conditional flow model learns the distribution of lattice descriptors for a given $(\mathrm{SMILES},~H,~Z')$ input. It transforms samples from a simple noise distribution through a learned continuous evolution~\cite{Lipman2023}. Different initial noise samples yield alternative descriptor vectors for the same input, allowing the model to explore a range of plausible cells.

The flow model generates the 12-dimensional vector $\mathbf{x} = \bigl[S(6),\;\log\lambda(3),\;\log\lambda^\ast(3)\bigr]$, comprising 6 Selling scalars and the logarithms of the 3 direct- and 3 reciprocal-lattice successive minima. The model factorizes this vector into two conditional flows: a Selling flow samples $S$, and a second flow samples $\bigl[\log\lambda,\;\log\lambda^\ast\bigr]$ conditioned on the sampled $S$.

As shown in Fig.~\ref{fig:pipeline}a, a molecular graph is constructed from the SMILES string using \texttt{RDKit}~\cite{LandrumRDKit}. The 33-dimensional node features use one-hot encodings of element identity within a restricted organic element set, degree, formal charge, aromaticity, hybridisation, and implicit hydrogen count. Four-dimensional edge features encode bond type. A four-layer GINE encoder~\cite{Hu2020} with a hidden width of 256, followed by mean pooling and a linear projection, produces a graph embedding $h\in\mathbb{R}^{512}$. This embedding is concatenated with learned embeddings of the Hall index (64 dimensions) and $Z'$ (32 dimensions) to form the base condition

\begin{equation}
\mathbf{c}_0=\bigl[h\,\|\,\mathrm{emb}(H)\,\|\,\mathrm{emb}(Z')\bigr].
\end{equation}

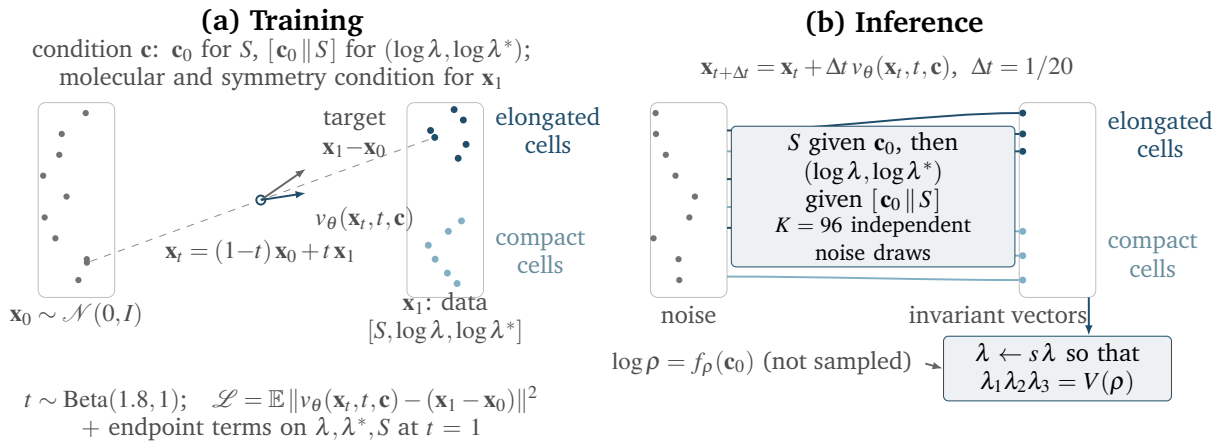
\begin{figure*}[htbp]
\centering
\resizebox{0.9\textwidth}{!}{%
\begin{tikzpicture}[
  font=\normalsize,
  arr/.style={-{Latex[length=1.5mm]},thick,accent},
  garr/.style={-{Latex[length=1.5mm]},thick,black!60},
  pt/.style={circle,fill=black!55,inner sep=0.9pt},
  d1/.style={circle,fill=bar1,inner sep=1.0pt},
  d2/.style={circle,fill=bar4,inner sep=1.0pt},
  nbox/.style={draw=accent,rounded corners=2pt,fill=accent!8,align=center,inner sep=2.5pt},
  note/.style={font=\normalsize,text=black!70,align=center}
]
\node[font=\bfseries\large,anchor=center,align=center] at (3.5,4.15) {(a) Training};
\draw[black!25,rounded corners=3pt] (0,0.2) rectangle (1.1,3.0);
\foreach \y in {0.45,0.75,1.05,1.35,1.65,1.95,2.25,2.55,2.85} {\node[pt] at (0.35+0.4*rand,\y) {};}
\node[note] at (0.55,-0.05) {$\mathbf{x}_0\sim\mathcal{N}(0,I)$};
\draw[black!25,rounded corners=3pt] (5.3,0.2) rectangle (6.4,3.0);
\foreach \y in {2.2,2.4,2.6,2.75,2.9} {\node[d1] at (5.85+0.3*rand,\y) {};}
\foreach \y in {0.4,0.55,0.75,0.95,1.15,1.3} {\node[d2] at (5.85+0.3*rand,\y) {};}
\node[note] at (5.85,-0.10) {$\mathbf{x}_1$: data\\$[S,\log\lambda,\log\lambda^\ast]$};
\node[note,anchor=west,text=bar1] at (6.45,2.6) {elongated\\cells};
\node[note,anchor=west,text=bar4!80!black] at (6.45,0.85) {compact\\cells};
\node[pt] (x0) at (0.7,0.7) {};
\node[d1] (x1) at (5.7,2.5) {};
\draw[black!45,dashed] (x0) -- (x1);
\coordinate (xt) at ($(x0)!0.5!(x1)$);
\node[circle,fill=white,draw=accent,thick,inner sep=1.3pt] at (xt) {};
\node[note,anchor=north] at (3.2,1.15) {$\mathbf{x}_t=(1{-}t)\,\mathbf{x}_0+t\,\mathbf{x}_1$};
\draw[garr] (xt) -- ++(0.65,0.45) node[note,anchor=west,xshift=3pt,yshift=14pt] {target\\$\mathbf{x}_1{-}\mathbf{x}_0$};
\draw[arr] (xt) -- ++(0.65,0.10) node[note,anchor=north west,xshift=0pt,yshift=-2pt] {$v_\theta(\mathbf{x}_t,t,\mathbf{c})$};
\node[note,text width=7.6cm] at (3.5,3.55) {condition $\mathbf{c}$: $\mathbf{c}_0$ for $S$, $[\,\mathbf{c}_0\,\|\,S\,]$ for $(\log\lambda,\log\lambda^\ast)$;\\molecular and symmetry condition for $\mathbf{x}_1$};
\node[note,text width=7.6cm,anchor=north] at (3.5,-0.95) {$t\sim\mathrm{Beta}(1.8,1)$;\quad $\mathcal{L}=\mathbb{E}\,\|v_\theta(\mathbf{x}_t,t,\mathbf{c})-(\mathbf{x}_1-\mathbf{x}_0)\|^2$\\$+$ endpoint terms on $\lambda,\lambda^\ast,S$ at $t=1$};
\begin{scope}[xshift=8.8cm]
\node[font=\bfseries\large,anchor=center,align=center] at (3.5,4.15) {(b) Inference};
\draw[black!25,rounded corners=3pt] (0,0.2) rectangle (1.1,3.0);
\foreach \y in {0.45,0.75,1.05,1.35,1.65,1.95,2.25,2.55,2.85} {\node[pt] at (0.35+0.4*rand,\y) {};}
\node[note] at (0.55,-0.05) {noise};
\draw[black!25,rounded corners=3pt] (5.3,0.2) rectangle (6.4,3.0);
\node[note,anchor=east] at (6.4,-0.05) {invariant vectors};
\foreach \ys/\ye in {2.6/2.85, 1.9/2.55, 1.2/2.3} {\draw[bar1,thick] (1.1,\ys) .. controls (2.5,\ys) and (4.0,\ye) .. (5.3,\ye); \node[d1] at (5.35,\ye) {};}
\foreach \ys/\ye in {0.5/0.45, 1.5/0.8, 2.3/1.15} {\draw[bar4,thick] (1.1,\ys) .. controls (2.5,\ys) and (4.0,\ye) .. (5.3,\ye); \node[d2] at (5.35,\ye) {};}
\node[note] at (3.4,3.5) {$\mathbf{x}_{t+\Delta t}=\mathbf{x}_t+\Delta t\,v_\theta(\mathbf{x}_t,t,\mathbf{c})$, \ $\Delta t=1/20$};
\node[nbox,text width=3.9cm] (c) at (3.2,1.65) {$S$ given $\mathbf{c}_0$, then\\$(\log\lambda,\log\lambda^\ast)$ given $[\,\mathbf{c}_0\,\|\,S\,]$\\{\small $K=96$ independent\\noise draws}};
\node[note,anchor=west,text=bar1] at (6.45,2.55) {elongated\\cells};
\node[note,anchor=west,text=bar4!80!black] at (6.45,0.8) {compact\\cells};
\node[nbox,text width=3.1cm,anchor=north] (rs) at (5.85,-0.35) {$\lambda\leftarrow s\,\lambda$ so that\\$\lambda_1\lambda_2\lambda_3=V(\rho)$};
\draw[arr] (6.3,0.2) -- (6.3,-0.35);
\node[note,anchor=east] at (3.9,-0.75) {$\log\rho=f_\rho(\mathbf{c}_0)$ (not sampled)};
\draw[garr] (3.95,-0.75) -- (rs.west);
\end{scope}
\end{tikzpicture}}
\caption{Conditional flow model. (a) Training pairs a Gaussian sample $\mathbf{x}_0$ with a data vector $\mathbf{x}_1$ and its molecular and symmetry condition, draws $t\sim\mathrm{Beta}(1.8,1)$, and regresses the network velocity at the interpolated point onto the straight-line displacement $\mathbf{x}_1-\mathbf{x}_0$, given the condition $\mathbf{c}$. (b) At inference the learned field is integrated from noise with 20 Euler steps. Each draw samples the Selling scalars first and the minima conditioned on them; different noise samples reach different modes of the conditional distribution (elongated versus compact cells). $K=96$ independent draws are taken per condition.}
\label{fig:flow}
\end{figure*}

The Selling flow is conditioned on $\mathbf{c}_0$, and the flow for the successive minima on $\mathbf{c}_0$ extended by the Selling scalars,
\begin{equation}
\mathbf{c}_\lambda=\bigl[\mathbf{c}_0\,\|\,S\bigr],
\end{equation}
so that each sampled metric receives its own direct and reciprocal minima. During training, the second flow is conditioned on the Selling scalars of the structure; at inference, on the sampled ones. Because the conditional distribution of lattice descriptors is multimodal, for example with compact and elongated cells for the same molecule and Hall setting, different noise samples reach different modes (Fig.~\ref{fig:flow}).

Conditional flow matching learns, for each of the two flows, a velocity field $v_\theta(\mathbf{x}_t,t,\mathbf{c})$ that transports Gaussian noise, $\mathbf{x}_0\sim\mathcal{N}(0,I)$, towards the distribution of lattice descriptors (Fig.~\ref{fig:flow}). For each training pair and each flow, we draw $t\sim\mathrm{Beta}(1.8,1)$, interpolate linearly between the noise sample $\mathbf{x}_0$ and the data vector $\mathbf{x}_1$, and train the predicted velocity to match the displacement $\mathbf{x}_1-\mathbf{x}_0$. At inference, the resulting ordinary differential equation is integrated using 20 Euler steps. Each velocity field is parameterised by a multilayer perceptron with a sinusoidal time embedding.

Auxiliary endpoint losses on the physical $\log\lambda$ and $\log\lambda^\ast$ values, a relative-$\lambda$ term, and a $\lambda$-product term constrain the axis ratios and the product $\lambda_1\lambda_2\lambda_3$. A Selling endpoint loss regularises the reduced metric. An optional axis-permutation head ranks the length permutations permitted by orthorhombic settings; it is not used for monoclinic cells.

The model was trained with AdamW~\cite{loshchilov2018decoupled} (learning rate $3\times10^{-4}$, weight decay $10^{-3}$, cosine schedule) and a batch size of 128 for 400 epochs, keeping the checkpoint with the lowest validation loss. 
Density $\rho$ is predicted deterministically by a separate head and provides an additional target for cell reconstruction.

\subsubsection{Cell reconstruction and selection}

Next, the sampled lattice descriptors are converted into explicit cell parameters as shown in Fig.~\ref{fig:pipeline}b. Here, we sample $K=96$ cell vectors, each from an independent noise draw. Vectors with extreme ratios between successive minima are discarded before reconstruction.

For each retained vector, lattice QRS searches the free cell parameters permitted by the specified Hall setting. The objective combines equally weighted terms measuring agreement with the predicted $\lambda$, $\lambda^\ast$, Selling scalars, and density $\rho$. The initial search box is constructed from the predicted successive minima. For monoclinic cells, $\lambda_1$ is assigned to the unique axis $b$, with the remaining minima assigned to $a$ and $c$. The search proceeds through 12 Sobol stages\cite{Sobol1967} of 1024 samples each. After each stage, the 32 best candidates are retained, and the search box is contracted around them by a factor of 0.4, subject to a minimum span. According to our test, reducing the search to 6 stages increases the median edge-length error against oracle targets from 0.3\% to 2.3\%, supporting the use of 12 stages.

After the removal of duplicates, 64 cells are selected for molecular packing QRS. Selection balances agreement with the predicted descriptors against diversity in cell shape and uses no experimental cells, matcher scores, or energies. Four ranked channels allocate slots to QRS-derived cells based on $\lambda$ consistency (6 slots), lattice-QRS loss (6), density and volume (4), and disagreement between the QRS cell and a direct Selling reconstruction (4). The remaining slots are filled by farthest-point selection in a standardized feature space of $\log\lambda$ ratios and cell angles.

For monoclinic cells, an additional channel reserves 12 of the 64 slots for alternative unique-axis assignments. These slots are taken from the shape-diversity allocation. Each cell is assigned to the first channel that selects it, preventing duplicate selection across channels; unfilled slots in any ranked channel are reassigned to the diversity channel.

\subsubsection{Molecular packing sampling}

For each selected cell, we search for molecular packing using the QRS algorithm~\cite{ZhuHattori2025} implemented in \texttt{PyXtal}~\cite{Fredericks2021}. As described in Fig.\ref{fig:pipeline}c, a conformer pool is generated once from up to 200 torsion samples, filtered by root-mean-square deviation (RMSD), and reused across generations. Within each generation, Sobol points sample the remaining packing degrees of freedom. For a rigid molecule with $Z'=1$, these comprise the fractional coordinates of its center and three Euler angles for the orientation. The sampled configurations are then mapped onto a discrete grid.

Each candidate is relaxed in CHARMM~\cite{Brooks2009} using an OpenFF force field~\cite{Boothroyd2023}. The primary search pass relaxes both molecular coordinates and lattice parameters, after which the structures are ranked by energy. The search uses up to $N_{\mathrm{gen}}=20$ generations, a population size of $N_{\mathrm{pop}}=96$ (typically done within 48 parallel CPU cores). In the benchmark, an individual cell search terminates after more than three experimental matches are identified, but the sweep continues until all 64 cells have been searched. 

\subsubsection{Evaluation}
To evaluate the efficiency of this approach, we compute success rate of sampling on organic systems selected from our recent work \cite{ZhuHattori2025}. A system is eligible if it has a single-component SMILES string, a Hall setting represented in the model vocabulary, and a pipeline value of $Z'\le1$ after the special-position step. These criteria retain 84 systems as shown in Table~\ref{tab:eligible}.

\begin{table}[htbp]
\centering
\caption{Benchmark eligibility and final structural coverage. A system is covered if at least one packing search produces a match to the experimental structure.}
\label{tab:eligible}
\setlength{\tabcolsep}{5pt}%
\begin{tabular}{@{}lrl@{}}
\toprule
 & $N$ & Codes \\
\midrule
Total systems~\cite{ZhuHattori2025} & 100 & \\
Eligible ($Z'\le 1$) & 84 & \\
Covered, lattice free & 80 & \\
Covered, frozen pass only & 3 & OBEQUJ, XAFPAY, \\
 & & XULDUD01 \\
Zero hits in both passes & 1 & QAXMEH53 \\
\midrule
Coverage & \textbf{83/84} & \textbf{98.8\%} \\
\bottomrule
\end{tabular}
\end{table}

A system is counted as covered if at least one flow generated cell searches produces a match to the experimental structure. If the entire primary sweep yields no matches, the same cells are searched again with the lattice frozen mode (Section~\ref{sec:adaptive}). A match from the lattice frozen mode also counts towards coverage. Otherwise, the system is classified as a miss. 

In structural match, we compare all structures with the experimental reference structure using the  \texttt{StructureMatcher} within the pymatgen~\cite{Ong2013}. The matching criteria use \texttt{ltol}=\texttt{stol}=0.3, an angle tolerance of $5^\circ$, and a site-distance cutoff of 0.3\,\AA, after the removal of H atoms. The success rate (SR) for an individual cell is the percentage of its relaxed structures that match the experimental reference. SR is used only for evaluation, not ranking. 

\section{Results}\label{sec:results}

\subsection{\texorpdfstring{Coverage on $Z'\le 1$ organics}{Coverage on Zprime <= 1 organics}}

Among all 84 eligible systems, we found that 80 matched with the lattice free. Table~\ref{tab:hard} shows that the coverage criterion is met with very different margins. At one extreme, many successes are redundant: ADAMAN01 matches in 63 of 64 cells, NAPHTA15 in 61, SITJUC in 59, and UREAXX02 in 56, so many handful of proposed cells would have sufficed with high SR values. 

\begin{table}[htbp]
\centering
\caption{Coverage statistics among selected examples.}
\label{tab:hard}
\begin{tabular}{@{}lrr@{}}
\toprule
CSD code & Hits & Max.\ SR (\%) \\
\midrule
ADAMAN01 & 63/64 & 81.3 \\
NAPHTA15 & 61/64 & 79.2 \\
SITJUC & 59/64 & 16.7 \\
UREAXX02 & 56/64 & 55.2 \\
$\cdots$ & $\cdots$ & $\cdots$ \\
OBEQIX & 6/64 & 0.40 \\
DOCXEA & 3/64 & 0.63 \\
MERRAF & 3/64 & 44.8 \\
XAFPAY03 & 1/64 & 0.05 \\
UJIRIO05 & 1/64 & 0.05 \\
\bottomrule
\end{tabular}
\end{table}

On the other hand, Table~\ref{tab:hard} also shows a list of systems succeed in only a few cells and usually at very low success rates. OBEQIX matches in six cells with a maximum SR of 0.40\%, while UJIRIO05 and XAFPAY03 each match in a single cell at 0.05\%. These systems satisfy the coverage criterion, but they generally reflect that it is harder to reproduce the target structures for these systems. These results are roughly consistent with our previous findings using the stochastic approaches~\cite{ZhuHattori2025}, where cases like OBEQIX (0.0009\%), XAFPAY03 (0.0013\%) require over 128k random samples. For these cases, our QRS approach is determinant and hence becomes more reliable in identifying the target results as compared to the stochastic sampling. In addition, MERRAF illustrates a different failure mode. Only three cells produce matches, but the best of them reaches an SR of 44.8\%: once the right cell is proposed the packing is easy, and the difficulty lies entirely in proposing it.

\subsection{Failure analysis}
Apart from the 80 out of 84 successes, it is important to understand why the rest 4 systems (QAXMEH53, XULDUD01, OBEQUJ, XAFPAY) failing to recover an experimental structure with the lattice free. We relaxed their CSD structures directly using the same OpenFF/CHARMM protocol (allowing the lattice to vary), and then compared the relaxed structures with the experimental references using the benchmark matching criteria. This control bypasses cell generation and packing sampling, testing whether the relaxation protocol preserves a known target structure.

\begin{figure}
    \centering
    \includegraphics[width=\columnwidth]{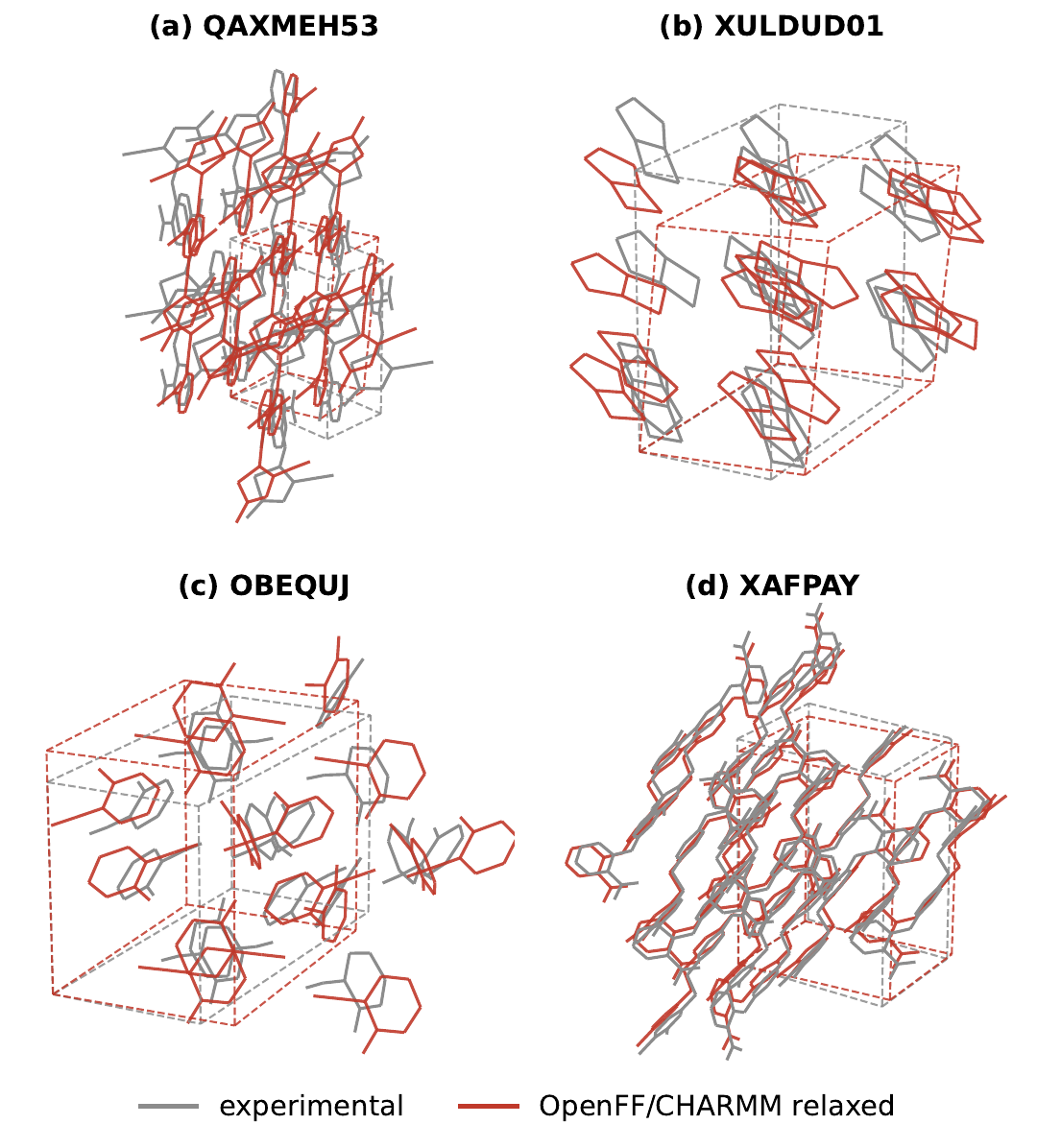}
    \caption{Overlays of experimental structures (grey) and structures after OpenFF/CHARMM relaxation (red): (a) QAXMEH53, (b) XULDUD01, (c) OBEQUJ, and (d) XAFPAY. Dashed lines indicate the corresponding unit cells; H atoms are omitted for clarity.}
    \label{fig:relax-overlays}
\end{figure}

Fig.~\ref{fig:relax-overlays} shows the graphical outcome. For QAXMEH53, XULDUD01, and OBEQUJ, relaxation moves the reference structure outside the matching criteria. QAXMEH53 undergoes substantial shear, with $\Delta\beta\approx-20^\circ$ and $\Delta\gamma\approx+16^\circ$. For XULDUD01, $a$ increases from 4.95 to 6.19\,\AA\ and $\beta$ decreases from $90.6^\circ$ to $80.8^\circ$. The cell lengths of OBEQUJ change from $7.38\times16.19\times9.65$\,\AA\ to $8.77\times13.10\times11.04$\,\AA, with a volume increase of 116\,\AA$^3$. The evidence for XAFPAY is less conclusive: the reported RMSD is 0.15\,\AA, despite changes of $\Delta\beta\approx-4^\circ$ and $\Delta V\approx-95$\,\AA$^3$. On the other hand, XAFPAY seems to well preserved after OpenFF/CHARMM relaxation. As long as the structure is identified from the search, the structure can be matched with a bit loosen criteria. 

\subsection{Adaptive lattice relaxation}\label{sec:adaptive}

The relaxation controls motivate a two-pass search protocol. We first search all 64 proposed cells with the lattice free to relax. If this sweep produces no matches, we repeat the packing searches once using the same proposed cells with their lattice parameters fixed. This fallback retains the learned cell geometry while allowing the molecular packing to relax. 

The frozen-lattice searches recover OBEQUJ, XULDUD01, and XAFPAY, increasing coverage from 80 to 83 of the 84 systems (Table~\ref{tab:eligible}). OBEQUJ matches in four cells, with an RMSD as low as 0.12\,\AA, while XULDUD01 matches in five cells, four of them in the first generation, with a maximum SR of 24.0\%. For OBEQUJ, the closest proposed cell lies within 4\% of the experimental Niggli-reduced cell lengths and angles, yet none of the 64 free-lattice searches produced a match. Recovery with a fixed lattice, together with the direct-relaxation control, supports the conclusion that lattice relaxation can prevent recovery even when a suitable cell has been proposed. XAFPAY is a marginal case: a single structure in one cell matches (SR 0.05\%), with an RMSD of 0.15\,\AA\ and a maximum site deviation of 0.24\,\AA\ against the 0.3\,\AA\ cutoff. Matches at such low SR are therefore sensitive to the software environment and choices of threshold values for claiming a structure match. Last, QAXMEH53 remains unmatched in both passes, showing that the fallback does not resolve all failures.

Because the second pass is restricted to systems with no primary match, it preserves the successes of the completed free-lattice searches and adds computation only for four of the 84 systems (4.8\%). The trigger nevertheless depends on comparison with the experimental reference, so this adaptive protocol is a benchmark procedure. A prospective CSP calculation would need to run both passes or use a separately validated criterion that does not require the reference structure.

\subsection{Computational Cost}
In a standard CSP run, lattice flow sampling and lattice QRS usually finish in minutes. The major cost is from conformational QRS and mostly depends on chemistry and number of atoms in the unit cell. As shown in Table~\ref{tab:cost}, a four-heavy-atom rigid crystal (UREAXX02) takes one to two minutes per cell, while 40-heavy-atom polyaromatic hydrocarbons quaterrylene (QUATER10) or a flexible molecule with a large conformer pool (e.g., C8-BTBT YOKBIK~\cite{Izawa2008}, 170 conformers) takes 10--20 minutes per cell. The rubrene QQQCIG04 has both a large number of atoms and conformational flexibility, taking 12--21 minutes per cell. These times also depend on the CPU type: the same search can run up to about twice as fast on AMD EPYC (Genoa, Turin) nodes as on Intel Xeon (Cascade Lake) nodes.
A 64-cell sweep with up to 20 generations per cell therefore requires hours of wall time on the stated hardware. These timings describe structure generation and force-field relaxation; they do not include the higher-level ranking needed for a full CSP task~\cite{Reilly2016,Hunnisett2024,ZhuHattori2025}.

\begin{table}[htbp]
\centering
\caption{Conformational-QRS cost per cell on 48 CPUs. Ranges span the median time per cell across benchmark campaigns run on different nodes; the time depends on the CPU type, with AMD EPYC (Genoa, Turin) nodes up to about twice as fast as Intel Xeon (Cascade Lake) nodes.}
\label{tab:cost}
\begin{tabular}{@{}lll@{}}
\toprule
Pattern & Example & Time / Cell \\
\midrule
Small rigid & UREAXX02 & 1--2 min \\
Large rigid PAH & QUATER10 & 11--15 min \\
Flexible, 16 conformers & MIVDEC & 15--16 min \\
Large conformer pool & YOKBIK & 10--20 min \\
Large and flexible & QQQCIG04 & 12--21 min \\
\bottomrule
\end{tabular}
\end{table}

\section{Discussion}\label{sec:discussion}
The closest comparison is the population-based CSP workflow our earlier work \cite{ZhuHattori2025}, which samples unit cells at random and evolves a population of complete crystals jointly exploring cell geometry and molecular packing. In CellNet, unit cells are drawn from a learned conditional distribution, and each proposed cell undergoes a quasi-random packing search with a fixed Sobol sequence. This approach offers two advantages. First, fixed settings and seeds make the cell proposals and packing searches reproducible within a given software environment, enabling controlled comparisons of protocol changes using identical flow draws. Second, the computational budget is bounded by the number of cells and the per-cell packing budget (Table~\ref{tab:cost}), rather than by the number of generations needed for a population to converge. This decomposition makes coverage dependent on the quality of the proposals: recovery becomes difficult when the target cell lies far from the proposed cells. However, our benchmark suggests that the proposals are generally adequate for all $Z'\le1$ organic systems.

To demonstrate the proof of concept, we only assess structural coverage for a supplied Hall setting and pipeline $Z'\le1$. Hence, they do not establish polymorph stability rankings or performance across all space groups or larger asymmetric units. The relaxation controls show that the energy model can limit recovery even when a suitable cell is available, by driving structures away from the experimental packing. Fixed-lattice searches mitigate this limitation for OBEQUJ, XULDUD01, and, marginally, XAFPAY, but QAXMEH53 remains uncovered. The reported coverage therefore reflects the combined performance of cell generation, packing search, and force-field relaxation. Moreover, the benchmark uses experimental matches to terminate individual searches and trigger the frozen-lattice fallback; prospective applications will require criteria that do not depend on the reference structure, together with more accurate energy ranking. Extending the flow to sample the Hall setting and $Z'$ as well as the cell is a natural next step towards fully blind prediction.

\section{Conclusions}\label{sec:conclusion}

In this work, we propose CellNet to combine learned unit-cell proposals with symmetry-aware molecular packing searches for crystal structure prediciton. Given a molecular graph, a Hall setting, and $Z'$, the conditional flow model generates lattice descriptors from which compatible cells are reconstructed by quasi-random sampling. A subsequent search explores molecular packings within each proposed cell. On the single-component benchmark with pipeline $Z'\le1$, this approach recovers 83 of 84 experimental structures (98.8\%): 80 with lattice relaxation and three through a frozen-lattice fallback triggered by the absence of experimental matches. These findings support the use of learned cell proposals to guide physics-based CSP searches. Further development will focus on search control without experimental references, more accurate energy models, and extension to unknown symmetry settings and larger asymmetric units.

\section*{Data availability}

SPaDe-CSP training tables are those of Taniguchi and Fukasawa~\cite{TaniguchiFukasawa2025}; the CellNet extract and precomputed lattice targets are stored under \texttt{datasets/spade-csp/} in the CellNet repository. The coverage benchmark uses \texttt{datasets/test.db} from Zhu and Hattori~\cite{ZhuHattori2025}. Experimental reference structures originate from the CSD~\cite{Groom2016} and are subject to the CCDC licence. 

\section*{Code availability}
The CellNet project is available at \url{https://github.com/qzhu2017/CellNet-prototype}.


\section*{Conflicts of interest}
There are no conflicts to declare.

\section*{Acknowledgments} 
Q.Z. acknowledges financial support from the NSF (DMR-2410178) and computing resources provided by ACCESS (TG-MAT230046).

\textbf{\large{References}}

\bibliography{rsc.bib}

\end{document}